\documentclass[12pt]{article}
\usepackage{epsfig}

\newcommand{\be}{\begin{eqnarray}}
\newcommand{\ee}{\end{eqnarray}}
\newcommand{\sll}{\raise.15ex\hbox{$/$}\kern-.43em\hbox{$l$}}
\newcommand{\slp}{\raise.15ex\hbox{$/$}\kern-.43em\hbox{$p$}}
\newcommand{\slq}{\raise.15ex\hbox{$/$}\kern-.43em\hbox{$q$}}
\newcommand{\slk}{\raise.15ex\hbox{$/$}\kern-.43em\hbox{$k$}}
\newcommand{\slepsilon}{\raise.15ex\hbox{$/$}\kern-.53em\hbox{$\epsilon$}}

\begin{document}

\bibliographystyle{unsrt}
\footskip 1.0cm

\thispagestyle{empty}
\begin{flushright}
INT--PUB 04--02
\end{flushright}
\vspace{0.3in}

\begin{center}{\Large \bf {Electromagnetic Signatures of the Color Glass 
Condensate: Dileptons}}\\

\vspace{1.8in}
{\large  Jamal Jalilian-Marian}\\

\vspace{.2in}
{\it Institute for Nuclear Theory, University of Washington, 
Seattle, WA 98195\\}

\end{center}

\vspace*{25mm}

\begin{abstract}

\noindent We evaluate the invariant cross section for production of 
dileptons in forward rapidities at RHIC and LHC, using the Color
Glass Condensate formalism and present results for the nuclear 
modification factor $R_{d(p)A}$ as a function of dilepton invariant mass
for the most central deuteron (proton)-nucleus collisions. 

\end{abstract}
\newpage

\section{Introduction}

The recent results on suppression of hadron spectra in deuteron gold (dA) 
collisions in the forward rapidity region \cite{brahms} of the Relativistic 
Heavy Ion 
Collider (RHIC) at the Brookhaven National Lab (BNL) has
generated a lot of interest in the applications of semi-classical QCD and the 
Color Glass Condensate \cite{cgc} to RHIC. Even though the Color Glass 
Condensate is the prediction of QCD for the wave function of a hadron or 
nucleus at high energies \cite{glr,mq,qcd-sum}, it is not a priori clear 
at what 
energy this happens. There is some experimental evidence \cite{kln} that 
RHIC may be at just high enough energy to see glimpses of the Color Glass 
Condensate. 

The applications of the Color 
Glass Condensate formalism to the heavy ion collisions at RHIC have been most 
successful at low $p_t$ \cite{kln} which probe the kinematic region where 
$x_{bj} \sim 0.01$ at mid rapidity. However, the mid rapidity region in
heavy ion collisions is not the best 
place to look for the Color Glass Condensate because of the dominance of 
the final state effects, such as the energy loss of energetic partons 
\cite{eloss} from the possibly formed Quark Gluon Plasma.

The forward rapidity region in dA collisions (the deuteron fragmentation 
region) is the best place in a hadronic/nuclear collision to probe the 
Color Glass Condensate \cite{djm}. First, there is presumably no Quark 
Gluon Plasma formed 
in a deuteron gold collision so that the dominant final state effects such 
the jet energy loss from the plasma are absent. Second, the forward rapidity
region probes the small $x_{bj}$ part of the nuclear wave function and 
the large $x_{bj}$ part of the deuteron wave function. This is the ideal 
situation for the 
Color Glass Condensate probes since the high gluon density effects in the 
nucleus which give rise to the Color Glass Condensate are
the strongest in this kinematics.

RHIC is a unique experiment in the sense that it has almost a continuous 
rapidity coverage, $0 < y < 4$,  among its various detectors where it can 
detect various particles. The STAR detector can measure hadrons and 
photons in mid rapidity as well as at  $y=4$. The PHENIX collaboration can 
measure hadrons at mid rapidity as well as dimuons in the rapidity region 
between $1.2 - 2.2$ while BRAHMS has measured hadrons at mid rapidity as 
well as rapidities of $1, 2.2$ and $3.2$. Therefore, one has the chance to 
map out the rapidity ($x_{bj}$) dependence of particle production and confront 
it with the predictions of the Color Glass Condensate formalism. Already, 
the qualitative agreement between the predictions of the Color Glass 
Condensate formalism \cite{kkt,aaksw,jmnr} and the data from BRAHMS, both the 
suppression of $R_{dA}$ and its centrality dependence, are quite remarkable 
specially since all the available models in the market missed this 
suppression despite their many free parameters \cite{av}.

Electromagnetic probes such as photons and dileptons provide another
tool, in addition to hadrons, with which to investigate the properties 
of the Color Glass Condensate \cite{gjm}. They are cleaner in the sense 
that they 
do not undergo strong interactions with the other partons produced
after the collision. Furthermore, in the case of hadrons, one needs to
convolute the produced parton with the desired hadron fragmentation
function. This means that a hadron measured at a given transverse momentum
$p_t$ comes from the fragmentation of a parton at a yet higher transverse
momentum $k_t = p_t/z$. Since the collision energy is fixed, this 
takes one to higher momenta which correspond to higher $x_{bj}$ where 
the Color Glass Condensate effects become less dominant. Dileptons and photons
do not suffer from this and are therefore, a better signature of the high
gluon density effects.

In this short paper, we provide a numerical analysis of the dilepton production
cross section in deuteron-gold collisions \cite{gjm} at RHIC and 
proton-lead collisions
at LHC in the forward rapidity region. We consider $y=2.2$ at RHIC where 
the PHENIX detector is located and $y=5$ at LHC where there are plans to
measure dileptons. We show our results for the absolute cross sections as well
as the nuclear modification factor $R_{d(p)A}$ and show that, similar to
hadrons, dilepton production is also suppressed in the forward rapidity region.

\section
{Dilepton Production in d(p)A Collisions}

\begin{figure}[htp]
\centering
\setlength{\epsfxsize=14cm}
\centerline{\epsffile{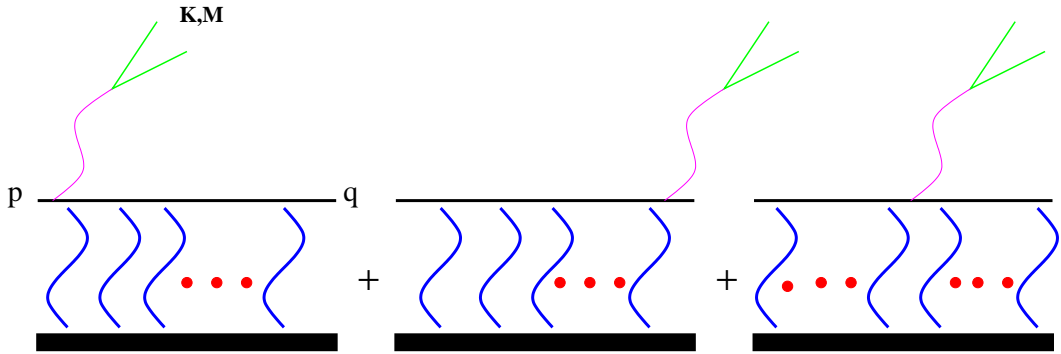}}
\caption{Dilepton production in quark-nucleus scattering.}
\label{fig:di_fig}
\end{figure}

Our starting point is the dilepton production in quark-nucleus scattering
using the Color Glass Condensate formalism as derived in \cite{gjm} (see
also \cite{boris} for an equivalent approach). The
diagrams corresponding to this process are shown in Fig. (\ref{fig:di_fig})
where the virtual photon is emitted before, after or during the multiple 
scatterings of the quark from the target nucleus. 

It can be shown that the diagram where the virtual photon is emitted 
during the scattering of the quark from the nucleus is suppressed by 
the Lorentz factor $\gamma$ and does not contribute. Assuming the 
incoming quark is on shell and ignoring all the quark masses, one gets 

\be
&&
{d\sigma^{q\,A\rightarrow q\,l^+l^-\,X}
\over dz\,d^2b_t\,d^2k_t\,d\log M^2}=
{{2 \alpha^2_{em}}\over{3\pi}}
\int \,{{d^2 l_t}\over{(2\pi)^4}} \,\,
\hat{\sigma}_{dipole}(x,b_t,l_t)
\nonumber\\
&&
\nonumber \\
&&
\Bigg\{
\bigg[{{1+(1-z)^2}\over{z}}\bigg]
{{z^2 l_t^2}\over{[k_t^2+M^2(1-z)]
[(k_t - z l_t)^2 + M^2(1-z)]}}\nonumber\\
&&
\nonumber \\
&&
- z(1-z)\,M^2\Bigg[{1\over{[k_t^2 + M^2(1-z)]}}-{1\over{[(
k_t - z l_t)^2 + M^2(1-z)]}}\Bigg]^2
\Bigg\}
\ee
with $l_t=q_t + k_t$ and $q_t$ and $k_t$ are the transverse momenta of 
the outgoing quark and dilepton pair respectively while $z$ is the 
fraction of the energy of the incoming quark carried away by the 
virtual photon and $M$ is the dilepton pair invariant mass. $\hat{\sigma}$
is the Fourier transform of the dipole cross section given in 
(\ref{eq:cs_iim}).

To proceed further, we will integrate over the transverse momentum of the
dilepton pair. This can be done analytically and simplifies the expression
for the cross section considerably. The transverse momentum integrated
cross section is also more useful experimentally since the production
rate is quite small, at least at RHIC. Rewriting the dipole cross section
in the coordinate space, we get
   
\be
{d\sigma^{q\,A\rightarrow q\,l^+l^-\,X}
\over d^2b_t\,d M^2}=&&
{{\alpha^2_{em}}\over{3\pi^2}}\, 
\int \,dz \,{1-z \over z^3} \int \, dr_t^2 \, 
\sigma_{dipole}(x_g,b_t,r_t)
\nonumber\\
&&
\nonumber \\
&&
\Bigg[[1 + (1-z)^2] \,K_1^2[{\sqrt{1-z}\over z} M r_t] + 
2 (1-z) \, K_0^2[{\sqrt{1-z}\over z} M r_t]\Bigg]
\label{eq:main}
\ee

To relate this to deuteron (proton)-nucleus scattering, we need to
convolute (\ref{eq:main}) with the quark (and anti-quark) distributions
$q(x,M^2)$ in a deuteron (proton). As shown in \cite{boris}, this can be
written in terms of the deuteron (proton) structure function $F_2$

\be
{d\sigma^{d(p)\,A\rightarrow l^+l^-\,X}
\over d^2b_t\,d M^2\, dx_F}= &&
{{\alpha^2_{em}}\over{6\pi^2}}{1 \over x_q + x_g}\, 
\int_{x_q}^1 \,dz \,\int dr_t^2 \, {1-z \over z^2} \, F_2^{d(p)} (x_q/z) \, 
\sigma_{dipole}(x_g,b_t,r_t)
\nonumber\\
&&
\nonumber \\
&&
\Bigg[[1 + (1-z)^2] \,K_1^2[{\sqrt{1-z}\over z} M r_t] + 
2 (1-z) \, K_0^2[{\sqrt{1-z}\over z} M r_t]\Bigg]
\label{eq:dpA}
\ee
where 
\be
x_q= {1\over 2} \bigg[\sqrt{x_F^2 + 4 {M^2 \over s}} + x_F\bigg ]\nonumber \\
x_g=  {1\over 2} \bigg[\sqrt{x_F^2 + 4 {M^2 \over s}} - x_F\bigg ]
\ee
and \[ F_2^{d(p)}\equiv \sum_f x \, [q_f(x,M^2) + \bar{q}_f (x,M^2)]\] is the 
deuteron (proton) structure function with 
$x_F\equiv {M\over \sqrt{s}}[e^y - e^{-y}]$.
 Note that the sum over quark and
anti-quark flavors is different for protons and deuterons. Here, we will
use eq. (\ref{eq:dpA}) to calculate the dilepton production cross section
in deuteron (proton) nucleus collisions. In case of a proton projectile, 
we use the GRV98 parameterization \cite{grv98} of the structure function 
$F_2^p$. We note that, in this kinematic region, the incoming quark or 
anti-quark 
distributions in a proton are well known and there is very little 
difference between different parameterizations. For a deuteron projectile, 
we use the HKM parameterization \cite{hkm} of the $F_2^d$ structure function 
which includes shadowing. Again, since the projectile quarks and anti-quarks 
are in the large $x_{bj}$ region, the nuclear effects in the projectile
deuteron are not large except at very large $x_{bj} \sim 0.7 - 0.9$ where
shadowing can be a $20\% - 30\%$ effect.

The forward rapidity region in a $d(p)A$ collision probes small $x_{bj}$
gluons in the nucleus. Therefore, the target nucleus is treated as a 
Color Glass Condensate. To proceed further, we need to know the dipole
cross section $\sigma_{dipole}(x_g,b_t,r_t)$. The dipole cross section
satisfies the JIMWLK equation \cite{jimwlk}. This equation has recently 
been solved on a lattice in \cite{hw}. Alternatively, one can solve the
large $N_c$ limit of the JIMWLK equation, known as the BK equation \cite{bk}.
This has been done numerically by several authors \cite{solbk}. 

Alternatively, one can model the dipole cross section based on the known
properties of the solution to the non-linear evolution equation in various
limits. This has been done in \cite{iim} and a comparison to the HERA
data on structure functions has been performed. It is shown in \cite{iim}
that one can fit the HERA data in the kinematic region $Q^2 < 50 \,GeV^2$
and $x_{bj} < 0.01$. Furthermore, this ansatz has a simple form which 
includes the right anomalous dimension in the extended scaling region 
\cite{ks,iim-scaling}, unlike some previous parameterizations of the 
dipole cross section \cite{gbw}. Having the
right anomalous dimension is extremely important specially since the
recent data from BRAHMS at RHIC may indicate that we are in the region
corresponding to the linear evolution and the BFKL anomalous dimension.

In \cite{iim} parameterization, the dipole cross section has the 
following simple form
\be
\int d^2 b_t \, \sigma_{dipole}(x_g,b_t,r_t) \equiv 
2\pi R^2  \,\,{\cal N} \,(x_g, r_t Q_s)
\label{eq:cs_iim}
\ee
where 
\be
{\cal N} \,(x_g, r_t Q_s) = 1 - e^{-a \ln^2 b\,r_t Q_s}  \,\,\,\,\,\,\,\,\,\,
r_t Q_s > 2 
\nonumber
\ee
and 
\be
{\cal N} \,(x_g, r_t Q_s) = {\cal N}_0 \exp \Bigg\{ 2 \ln ({r_t Q_s \over 2})
\bigg [\gamma_s + {\ln 2/r_t Q_s \over \kappa \lambda \ln 1/x_g} \bigg] \Bigg\}
\,\,\,\,\,\,\,\,\,\,
r_t Q_s < 2 
\label{eq:cs_param}
\ee
The constants $a, b$ are determined by matching the solutions at $r_t Q_s =2$
while $\gamma_s = 0.63$ and $\kappa = 9.9$ are determined from LO BFKL.
The form of the saturation scale $Q_s^2$ is taken to be 
$Q_s^2 \equiv (x_0/x)^{\lambda} \, GeV^2$ with $x_0, \lambda, {\cal N}_0$
determined from fitting the HERA data on proton structure function $F_2$.
We refer the reader to \cite{iim} for details of the fit. In case of a
nucleus, we make the assumption that the saturation scale of the nucleus
is $Q_{sA}^2 \equiv A^{1/3} Q_{sp}^2$. We now have all the ingredients
necessary to evaluate the dilepton production cross section as given
by eq. (\ref{eq:dpA}). 

It is instructive to consider the kinematic regions
where a parameterization of the form used in \cite{iim} is appropriate.
This parameterization is valid in the saturation region, defined by scales
$M < Q_s$, and in the extended scaling region defined by $M < Q_{es}$ where
$Q_{es} \equiv Q_s^2/Q_0$. In the case of the proton, the initial scale
$Q_0$ is of the order of $\Lambda_{QCD}$ while in the case of nuclei, it
is $Q_0 = Q_s(x_0)$ where $\ln 1/x_0$ is the initial rapidity where 
high gluon density effects become important. This is usually taken to be
$x_0 \sim 0.05 -0.01$. It is important to notice that the actual value of the
extended scaling scale does not enter anywhere in our results, but that it
just defines how high in dilepton invariant mass $M$ we can go in
this formalism. 

To get a feeling for the saturation and extended scaling 
scales, we note that at $y=2.2$, the extended scaling scale of a proton is
$\sim 3$ GeV for $M \sim 3$ GeV. This is why, at RHIC, we show results
for dilepton invariant masses up to $M \sim 3$ GeV. For rapidity
$y\sim 5$ at LHC, the extended scaling scale goes up to $13$ GeV.  

\begin{figure}[htp]
\centering
\setlength{\epsfxsize=10cm}
\centerline{\epsffile{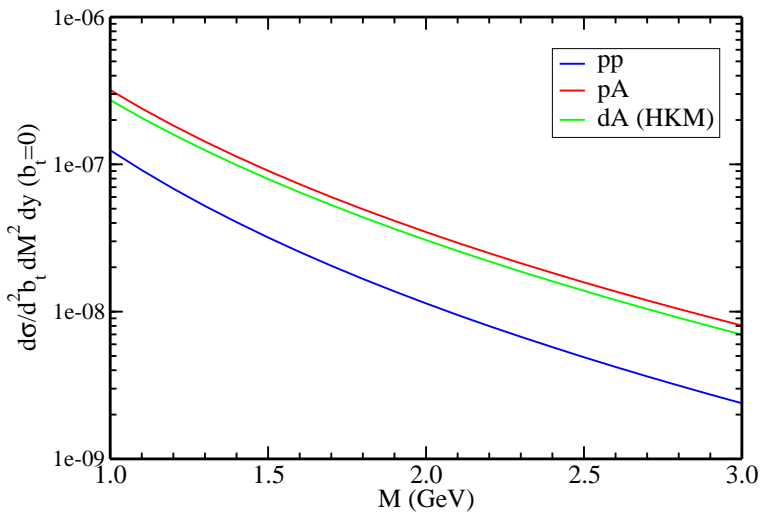}}
\caption{Dilepton production at RHIC: $y=2.2$ and $b_t = 0$.}
\label{fig:cs_rhic}
\end{figure}

We show our results for the absolute cross sections, as given in eq. 
(\ref{eq:dpA}), in 
Fig. (\ref{fig:cs_rhic}). For comparison, we show the cross sections for 
both proton-proton and deuteron-gold scattering for different 
dilepton invariant masses at rapidity $y=2.2$, appropriate for the 
PHENIX detector at RHIC. 

\vspace{0.3in}

It should be kept in mind that $M \sim 3$ GeV is quite likely at
the edge of the extended scaling region for this kinematics while
our formalism is best suited for the saturation and extended scaling 
region and could very well begin to break down as we go to higher dilepton
masses. Also, since the saturation and extended scaling scales are 
larger for a nucleus than a proton, our results are probably more
reliable for deuteron-gold collisions than proton-proton collisions
for this particular kinematics. This will improve as we go to higher
energies and/or higher rapidities such as those covered by LHC.

\begin{figure}[htp]
\centering
\setlength{\epsfxsize=10cm}
\centerline{\epsffile{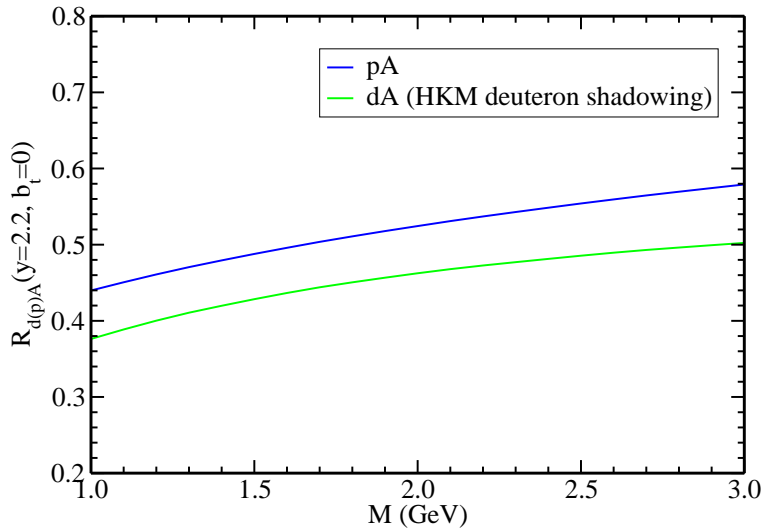}}
\caption{$R_{dA}$ and $R_{pA}$ at RHIC: $y=2.2$ and $b_t = 0$.}
\label{fig:R_rhic}
\end{figure}

The nuclear modification factor $R_{dA}$ defined as 
\be
R_{dA}\equiv {d\sigma^{dA \rightarrow l^+l^- X} /dy\, dM^2\,d^2b_t \over 
A^{1/3} d\sigma^{pp \rightarrow l^+l^- X} /dy\, dM^2\, d^2b_t}
\nonumber
\ee
is shown in Fig. (\ref{fig:R_rhic}) for the most central deuteron-gold
collisions. We also show the nuclear modification factor for the proton-gold
collisions for sake of comparison. It should be noted that nuclear
shadowing of the incoming deuteron wave function is taken into account in the 
HKM parameterization \cite{hkm}. As is clear from the figure, there is a 
large difference between dilepton production in proton-gold and 
deuteron-gold collisions as there must be since the flavor and 
as a result, electric charge, decomposition of proton and deuteron are 
different. 

\vspace{0.3in}

\begin{figure}[htp]
\centering
\setlength{\epsfxsize=10cm}
\centerline{\epsffile{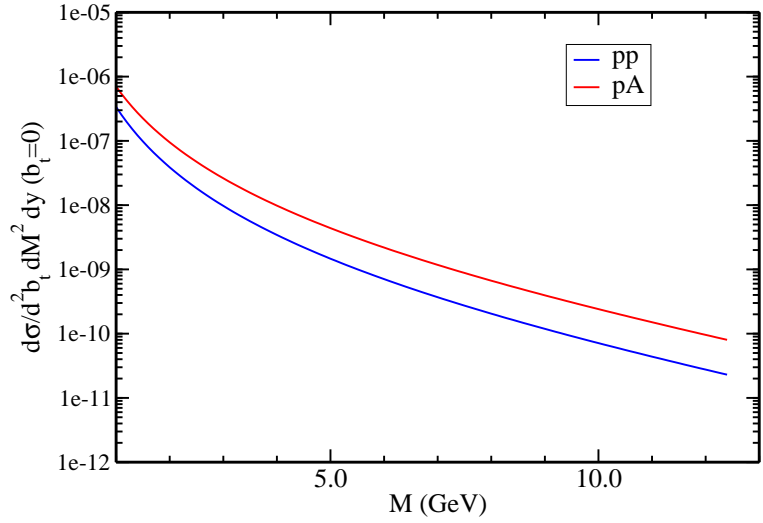}}
\caption{Dilepton production at LHC: $y=5$ and $b_t = 0$}
\label{fig:cs_lhc}
\end{figure}

In Fig. (\ref{fig:cs_lhc}), we show the dilepton production cross 
section in proton-proton and proton-lead collisions at LHC, at the 
rapidity of $y=5$. We now go to dilepton masses as high as $M \sim 13$ GeV
because the saturation scale is now much larger so that our formalism
is valid for higher dilepton invariant masses.

\vspace{0.3in}

\begin{figure}[htp]
\centering
\setlength{\epsfxsize=10cm}
\centerline{\epsffile{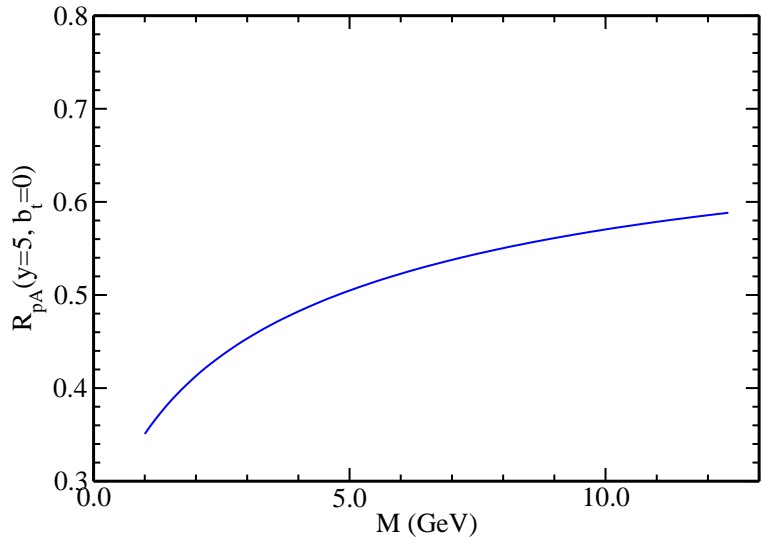}}
\caption{$R_{pA}$ at LHC: $y=5$ and $b_t =0$.}
\label{fig:R_lhc}
\end{figure}

In Fig. (\ref{fig:R_lhc}), we show our results for the nuclear modification 
factor $R_{pA}$ at LHC for the most central collisions proton-lead 
collisions. The suppression of the dilepton spectrum is much stronger 
than that at RHIC as expected since at LHC smaller $x_{bj}$ is probed. 

\vspace{0.3in}

\section{Discussion}

Semi-classical QCD extends the domain of applicability of weak 
coupling (perturbative) QCD to high energies where the naive
perturbative QCD approach fails due to high parton densities.
The resulting state of a hadron or nucleus at high energy where
parton densities are high is called a Color Glass Condensate. 
While the existence of such a high parton density state follows from  
QCD, the energy at which  this happens can not be derived,
at the moment, from the theory itself and needs to be determined 
experimentally. 

While there is some evidence in favor of Color Glass Condensate from 
HERA on electron-proton Deep Inelastic Scattering \cite{iim,gbw}, 
the BRAHMS collaboration at RHIC may have the best signature of the 
Color Glass Condensate in a nuclear environment so far in their  
measurement of the negatively charged hadrons in the forward rapidity 
region. To verify that this indeed the case, it is important to 
investigate the predictions of the Color Glass 
Condensate formalism for other processes such as dilepton production.

In this paper, we provide predictions for dilepton production cross sections
in proton-proton and deuteron (proton)-nucleus collisions in forward
rapidity regions at RHIC and LHC. The universal ingredient in this
cross section, as well as cross sections for hadronic observables, is the
quark anti-quark dipole- nucleus cross section which is subject to the
non-linear evolution equation and which has recently been solved. In this
work, we use an ansatz for the dipole cross section which has been 
successfully used to fit the HERA data on proton structure function.  
This model includes the physics of the BFKL anomalous dimension as well
as scaling properties of the Color Glass Condensate and therefore, is
well suited for our purpose. 

There are several caveats in this work which need a more careful 
treatment than considered here. First, the dipole ansatz as given 
in (\ref{eq:cs_param}) has been shown to work for a proton target, 
but has not been used for nuclei. In this work, we assumed that the 
$A$ dependence comes in through the saturation scale $Q^2_s A^{1/3}$. 
Ideally, one would like to have this only in the initial shape of the 
dipole. The subsequent $A$ dependence will then be 
determined by the non-linear equations. There are some indications that
as long as $x_{bj}$ is not too small, the assumed $A^{1/3}$ dependence
may be fine for large nuclei\cite{freund}. 

Another caveat is that the dipole model used here does not have the 
correct high $p_t$ behavior. In other words, it does not match onto the 
Double Log DGLAP limit which is contained in the Color Glass
Condensate. This may not be very important in the forward 
rapidity region since the $p_t$ coverage in this region is limited
due to kinematics of the experiment.

Most importantly, it remains to be seen whether dileptons at low invariant
mass (below $J/\psi$) peak can be measured at RHIC. This is most important
since this is where our predictions are most reliable at RHIC. If
and when RHIC future upgrades allow measurement of dileptons at higher
rapidities, one can go to higher dilepton masses since the saturation scale
is larger at more forward rapidities. Therefore, it is highly
desirable to have the capability to measure dileptons (as well as photons)
at RHIC in as forward rapidity as possible. Also, the STAR 
collaboration may be able to measure direct photons as well as photon 
$+$ jets at rapidity $y=4$ in the next deuteron-gold run at RHIC. The
photon $+$ jet process is unique in the sense it directly probes the
dipole cross section. Color Glass Condensate predictions for this process
will be presented elsewhere \cite{jjm}. Finally, having
another deuteron-gold run at RHIC in the {\it near future} will establish
conclusively whether the Color Glass Condensate has been observed at RHIC,
which seems to be the case, and will allow one to investigate its properties 
in detail.

\leftline{\bf Acknowledgments} 

We would like to thank R. Baier, F. Gelis, B. Kopeliovich, S. Kumano, 
A. Mueller and J. Raufeisen for useful discussions. This work is 
supported by DOE under grant number DOE/ER/41132.

\leftline{\bf References}

\renewenvironment{thebibliography}[1]
        {\begin{list}{[$\,$\arabic{enumi}$\,$]}  
        {\usecounter{enumi}\setlength{\parsep}{0pt}
         \setlength{\itemsep}{0pt}  \renewcommand{\baselinestretch}{1.2}
         \settowidth
        {\labelwidth}{#1 ~ ~}\sloppy}}{\end{list}}

\end{document}